\documentclass[aip, reprint]{revtex4-1}
\usepackage{graphicx}
\usepackage{siunitx}
\usepackage{color}
\usepackage{float}
\usepackage{amsmath}
\usepackage{xcolor}
\usepackage{soul}

\draft 

\begin{document}


\title{Confinement-controlled rectification in a geometric nanofluidic diode} 



\author{S. Dal Cengio}
\affiliation{Department of Condensed Matter, Universitat de Barcelona, Mart\'i i Franqu\'es 1 08028, Barcelona, Spain}

\author{I. Pagonabarraga}
\affiliation{Department of Condensed Matter, Universitat de Barcelona, Mart\'i i Franqu\'es 1 08028, Barcelona, Spain}
\affiliation{CECAM, Centre Europeen de Calcul Atomique et Mol\'eculaire, \'Ecole Polytechnique F\'ed\'erale de Lausanne,
Batochime, Avenue Forel 2, 1015 Lausanne, Switzerland}
\affiliation{Universitat de Barcelona Institute of Complex Systems (UBICS), Universitat de Barcelona, 08028, Barcelona, Spain}


\date{\today}

\begin{abstract}
Recent experiments with electrolytes driven through conical nanopores give evidence of strong rectified current response. In such devices, the asymmetry in the confinement is responsible of the non-Ohmic response, suggesting that the interplay of entropic and enthalpic forces plays a major role.
Here we propose a theoretical model to shed light on the physical mechanism underlying ionic current rectification (ICR). By use of an effective description of the ionic dynamics we explore the system's response in different electrostatic regimes.  
We show that the rectification efficiency, as well as the channel selectivity, is driven by the surface-to-bulk conductivity ratio Dukhin length rather than the electrical double layer overlap. 
\end{abstract}


\maketitle 

\section{INTRODUCTION}
Dating back to the famous thought experiment of Maxwell's demon (1867), the dream of designing \textit{force-free} transport devices has permeated different branches of physics, including nanofluidics.  In the context of nanofluidics, one can imagine the  \textit{ionic diode} \cite{B822554K}, a nanofluidic device  exhibiting ionic currents of unequal magnitude
under voltages of equal magnitude and opposite polarity, as a realization of such a demon. The first realization of such a nanometric ionic diode was reported by Siwy and Fuli\'nski in a geometrically asymmetric nanochannel obtained by asymmetric chemical
etching of a polymer foil \cite{PhysRevLett.89.198103}. Their conical channel demonstrated a strongly non linear ionic current under ac voltage, resulting in a net average current  under zero average forcing. Ionic current rectification (ICR) in conical nanochannels has since been extensively studied experimentally \cite{PhysRevLett.94.048102, doi:10.1021/nl061681k,  0957-4484-21-26-265301, doi:10.1021/nl504237k, Jubin4063}, thanks to the considerable progress made over the last twenty years in nano-fabrication technologies \cite{doi:10.1021/nn100692z}. ICR has also been observed in symmetric channels subject to a concentration gradient \cite{doi:10.1021/nl071770c} and in the presence of a surface charge discontinuity \cite{doi:10.1021/nl062806o}.\\ Empirically, the two features necessary to observe current rectification have been  identified as  the presence of surface charge and broken symmetry in the direction of transport, irrespectively of the nature of the broken symmetry. Alongside practical applications in macromolecular sensing and manipulation \cite{doi:10.1021/la061234k, NatureNanotec}, energy harvesting \cite{doi:10.1063/1.3001590, Siria2013, doi:10.1021/acsnano.6b00939} and water desalination \cite{doi:10.1021/acs.jpclett.7b01137, PhysRevLett.111.244501}, the phenomenon raises fundamental questions on the nature of ionic transport at the nanoscale. At this lengthscale, surfaces and entropic confinement strongly influence mass transport leading to the emergence of non-linear and exotic responses \cite{B909366B}, of which ICR is a prominent example. A rationalization of the latter would then be a testbed for understanding more complex behaviour occurring at the nanoscale \cite{B909366B} such as that of biological functionalized protein channels \cite{Sui2001, Tsong2002, doi:10.1021/jp111441p}.\\

In nano-sized fluidic diodes, electrostatic interactions between charged species play a key role. In the presence of a surface charge density $\sigma$ in contact with an electrolyte solution, an electrical double layer (EDL) builds up inside the channel with a characteristic decay length given by the Debye length, over which the imbalance of charge due to the channel walls is screened, 
\begin{equation}\label{eq:defDebye}
\lambda_D = \sqrt{\frac{ k_B T \epsilon_0 \epsilon_w}{c_{s} z^2 e^2} }.
\end{equation}
Here, $k_B$ is the Boltzmann constant, T is the temperature, $\epsilon_0$ and $\epsilon_w$ are respectively vacuum and relative water permittivities, $c_s$ is the bulk electrolyte concentration, $z$ is the electrolyte valency and $e$ is the elementary charge. At room temperature $\epsilon_w \approx 80$ and the Debye length can span from tens of nanometers down to a few \AA ngstroms depending on the salt concentration. Within the EDL an excess counterion concentration screens the surface charge giving rise to an electrically charged region. In the so-called \textit{entropic electrokinetic} regime \cite{10.3389/fphy.2013.00021} the Debye length is comparable to the tip of the nanopipette, \textit{i.e.}, the smallest aperture. This is typically the case in most synthetic realizations of nanochannels \cite{doi:10.1002/adfm.200500471, doi:10.1021/nn8007542, doi:10.1021/nl504237k} as well as in biological ion channels. Notably, measures of ICR in micrometer-sized systems have been reported more recently in the literature \cite{doi:10.1021/jacs.6b11696, doi:10.1021/acs.jpclett.7b03099}.
\\
It is convenient to introduce a second electrostatic length known as the Dukhin length: 
\begin{equation}\label{eq:defDukhin}
l_{Du} = \frac{|\sigma|}{e c_s} \sim  \frac{ \lambda_D^2}{l_{GC}} 
\end{equation}
 which quantifies the relative importance of  surface compared to bulk transport. Contrary to $\lambda_D$, the Dukhin length is a \textit{phenomenological} length: it does not directly correspond to a physically observable length in the system. Therefore it can be much larger or smaller than the system's size\cite{B909366B}. Eq.~(\ref{eq:defDukhin}) indicates that the   Dukhin length can be understood as the ratio between two different lengthscales, namely the Debye length (\ref{eq:defDebye}) and the Gouy-Chapman length $l_{GC} = 2 \epsilon_0 \epsilon_w k_B T/z^2 e |\sigma| $, defined as the typical length at which the surface electrostatic potential energy equals the thermal energy.
\\
 Nonetheless,  the electrostatic phase space, associated with the surface charge $\sigma$ and the bulk concentration $c_s$ degrees of freedom, is determined only by  two independent lengths. 
 As will become clear in the following, the choice of $\lambda_D$ and $l_{Du}$ as model parameters is convenient for the problem at hand.
 In analogy to colloidal science, we can also introduce a dimensionless Dukhin number \cite{19953-1}:
 \begin{equation}\label{eq:refDu}
 Du = \frac{l_{Du}}{\bar{h}}
 \end{equation}
 where $\bar{h}$ is the average half height of the channel.  $Du \gg 1$ identifies the  regime  globally dominated by surface transport.
\\
The theoretical literature on ICR has been confined mostly to numerical simulations of the ion dynamics using the classical Poisson-Nernst-Planck (PNP) equations for dilute electrolyte solutions \cite{PhysRevE.76.041202, doi:10.1021/jp911773m, doi:10.1021/jp111377h, C3CP51712H}. Such a framework  has quantitatively captured the phenomenon, demonstrating that a mean field continuum description is still valid for ionic dynamics down to a few nanometers. 
\\An early qualitative interpretation of ICR is traceable back to a paper of Dietrich Woermann \cite{B301021J} who rationalized the phenomenon in terms of ionic transference asymmetry between the ends of the channel. 
\\ At the same time, the study of particle transport over entropic barriers has attracted  the attention in non-equilibrium statistical physics \cite{10.3389/fphy.2013.00021, PhysRevE.83.051135, Yang9564, Marbach2018}. The first attempt to characterize transport in confined systems dates back to the early work of Jacobs \cite{jacobs1967diffusion} and Zwanzig \cite{doi:10.1021/j100189a004} who proposed the so-called \textit{Fick-Jacobs approach} (FJ) to account for the transport of Brownian particles geometrically confined in a quasi-one-dimensional system. Under the assumption of a separation of scales between the longitudinal and the transversal coordinates, the latter is integrated and the description is reduced to an effective 1D equation now containing an entropic term. The validity of the approach has been tested both in the case of free diffusion \cite{PhysRevE.75.051111, doi:10.1063/1.4934223} and in the presence of an external force \cite{PhysRevLett.96.130603} and demonstrated to be quantitatively accurate for channel geometries with smoothly varying cross-section under moderate external field, typically requirements that are satisfied in nanofluidic  setups{\cite{doi:10.1021/nn100692z, doi:10.1021/nl504237k, PhysRevLett.116.154501}} . \\ Overall the FJ approach represents a well-established systematic framework to describe transport in the presence of entropic barriers, and it has been recently extended to the regime of competition between energetic and entropic interactions in electrolyte dynamics \cite{PhysRevLett.113.128301}.
\\
Our goal in the present work is to gain insights on the fundamental mechanism controlling current rectification in a geometric diode, \textit{i.e.}, a conical channel with uniform charge density in contact with two reservoirs held at the same electrolyte concentration. Such a configuration corresponds to an extensively studied nanopipette experimental setup. Moreover, it represents  the conceptually intriguing case in which symmetry breaking originates only from the geometric confinement; such a system is thus able to harness \textit{entropy} to rectify ionic current. To address the problem we adapt the FJ approach to a 2D conical slab geometry. Contrary to previous works considering channels much larger than the Debye length \cite{PhysRevE.77.031131, Jubin4063}, the present formalism allows us to investigate the regime of finite $\lambda_D$ where partial Debye overlap occurs inside the channel, and to fully capture the interplay between energetic and entropic contributions. 
Furthermore, we are able to derive analytical predictions for the limiting conductance in the regime of strong EDL overlap which, to the best of our knowledge, have not yet been derived  for the geometric diode. Finally, our results assess the key role played by the Dukhin number  in the microscopic mechanism of rectification providing further insight on the nature of ICR. 
\\

\section{IONIC DYNAMICS}
As shown schematically in Fig. (\ref{FIG1}), we consider an open asymmetric channel with a slab geometry characterized by   longitudinal  size  $L$,  width $L_{z}$ and an x-dependent height
\begin{equation}
h(x) = \bar{h} + \frac{k L}{2} - k x
\end{equation}
where $\bar{h}$ is the half-aperture of the channel and $k = |d_x h| = (h_L - h_R)/L$ is the difference between the left $h_L$ and the right $h_R$ channel half-heights in units of channel length. In the following sections the channel slope is varied by keeping fixed its half-height $\bar{h}$ in order to compare systems with the same aspect ratio. 
\begin{figure}\label{FIG1}
\includegraphics[width=7cm]{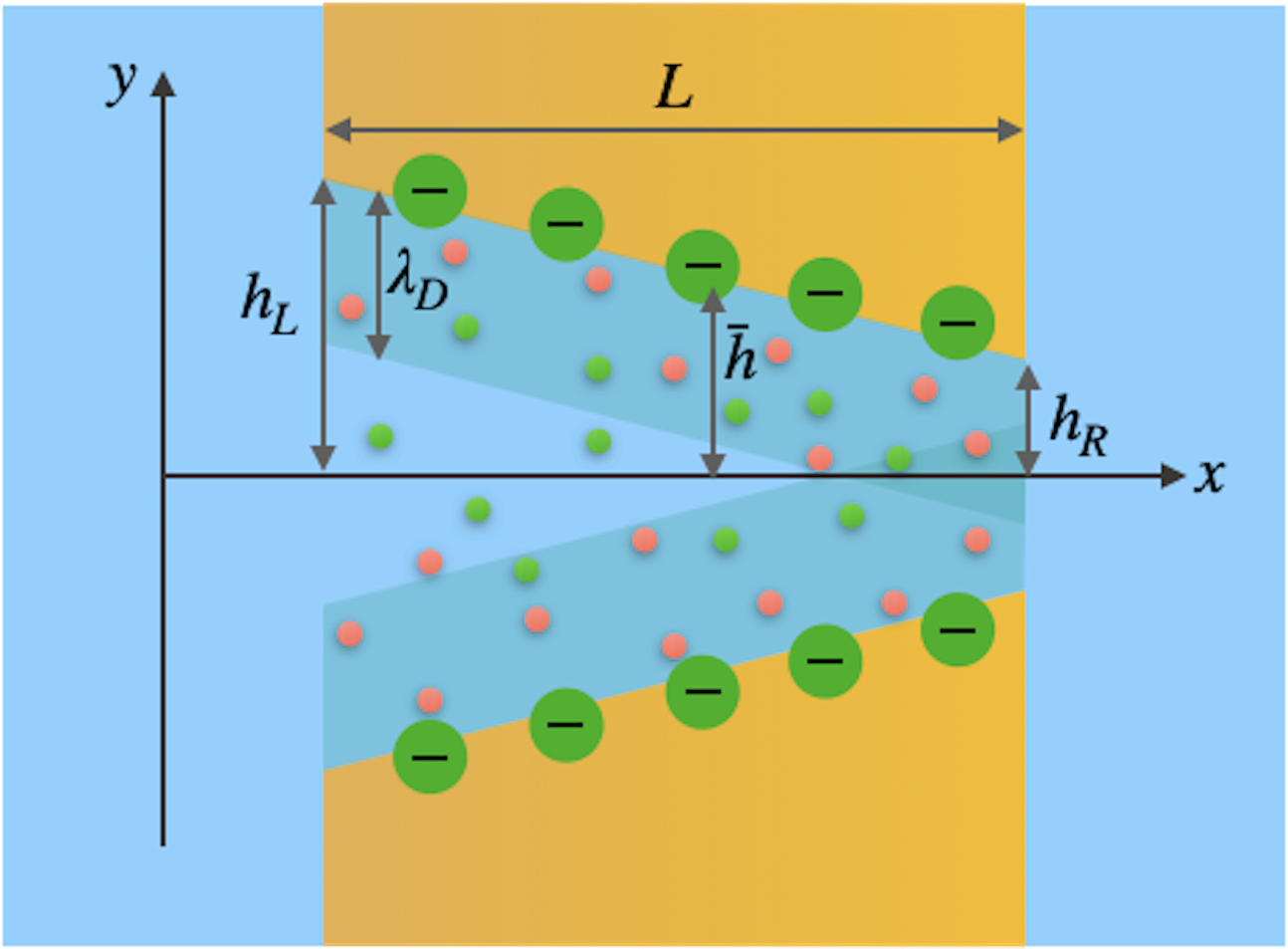}
\caption{\label{FIG1} Schematic view of the channel in contact with two reservoirs at fixed salt concentration. The channel width is assumed to be constant along the z direction pointing out of the page. The channel walls carry a uniform negative charge density and a  electrically charged double layer forms over a characteristic length $\lambda_D$.
}
\end{figure}
The channel is filled with a symmetric monovalent electrolyte composed of species having equal diffusion coefficient $D$, in contact with two reservoirs at fixed temperature $T$ and ionic strength $ c_s$. Each wall bears a uniform negative surface charge of density $\sigma < 0$. We assume $L_z \gg \bar{h}$ so that we can neglect the $z$ dependence of any variables of the model and the resulting  system is effectively 2D. 
\\
In order to characterize the ionic dynamics we derive effective one-dimensional transport equations for the ionic concentration profiles $c_{\pm}$. The approach relies on the constraint of a small aspect ratio $\epsilon = \bar{h}/L \ll 1$ , \textit{i.e.} a slowly-varying channel geometry. In this case, the transversal relaxation dynamics with characteristic relaxation time $\tau_y \sim \bar{h}^2/D$ is decoupled from the longitudinal relaxation dynamics with $\tau_x \sim L^2/D$ and the ions are assumed to \textit{instantaneously} adjust to the Boltzmann  distribution at each cross-section. Such separation of scale is known in the literature as local thermodynamic equilibrium \cite{degrootmazur} (LTE).
\\
Under these assumptions the steady-state Nernst-Planck equation for the positive and negative ionic species reads:
\begin{equation}\label{eqn:NP}
j_{\pm} = \mp D e \beta c_{\pm}(x, y) \partial_x \Phi(x,y) - D \partial_x c_{\pm}(x,y)
\end{equation}
where $j_{\pm}$ is the constant mass flux density along $x$, $\beta = 1/k_B T$  and $\Phi(x,y)$ is the total electrostatic potential inside the channel.
We have neglected in (\ref{eqn:NP}) the advective flux which proved to be minor compared to the electrophoretic contribution for moderate surface charge densities and moderate external fields \cite{doi:10.1021/jp911773m}.
\\
Eq.~(\ref{eqn:NP}) must be supplemented by the Poisson equation relating the electrostatic potential to the spatial charge distribution $q= e(c_+ - c_-)$ inside the channel:	
\begin{equation}\label{eq:Poisson}
\nabla^2 \Phi(x,y) = - \frac{q(x,y)}{\epsilon_o \epsilon_w}
\end{equation}
In the next section we reduce (\ref{eqn:NP}) to an effective 1D equation by introducing the FJ \textit{ansatz} for the ionic concentration profiles as explained in \ref{entropic}. For consistency, the same approximation is applied to the Poisson equation together with the assumption of small transversal variation of $\Phi$  (see section \ref{local}), which allows to formally integrate (\ref{eq:Poisson}).

\subsection{The Fick-Jacobs approach}\label{entropic}
Since the ionic transversal and longitudinal dynamics are assumed to be decoupled, it is convenient to introduce the marginal concentration as the cross-sectional integral of the volumetric concentration:
\begin{equation}
c_{\pm}(x) = \int_{-h(x)}^{+h(x)} dy\ c_{\pm} (x,y)
\end{equation}
Moreover, following the approach of Zwanzig \cite{doi:10.1021/j100189a004} we define  $x-$dependent free energies $A_{\pm}(x)$ via:
\begin{equation}
e^{-\beta A_{\pm}(x)} = \frac{1}{\bar{h}} \int_{-h(x)}^{+h(x)} dy\ e^{\mp \beta e \Phi(x,y)} 
\end{equation}
From the hypothesis of LTE we may factor the volumetric concentrations $c_{\pm}(x,y)$ into the product of equilibrium normalized conditional densities $\xi_{\pm}(y;x)$ and the marginal concentrations,
\begin{equation}\label{eq:FJhypo}
c_{\pm}(x,y) \approx \xi_{\pm}(y;x) \cdot c_{\pm}(x) = \frac{e^{\mp \beta e \Phi(x,y)}}{\displaystyle\int_{-h(x)}^{+h(x)} dy\ e^{\mp \beta e \Phi(x,y)}} \cdot c_{\pm}(x).
\end{equation}
Eq.~(\ref{eq:FJhypo}) represents the key \textit{ansatz} of the FJ approach. Martens \textit{et al} \cite{PhysRevE.83.051135} proved that (\ref{eq:FJhypo}) can be recovered as the zero-order term of a perturbative expansion in series for the geometrical parameter $k$ around the zero-transversal-flux solution. 
Notably for the case of a conical channel, where $|d_x h(x)| = \rm const $,  taking into account the extra $x-$dependence of the diffusivity $D(x)$ amounts to a rescaling of the diffusion coefficient thus making the theory here developed valid up to $k \le 1 $ \cite{doi:10.1063/1.4934223, doi:10.1063/1.4993129}.
\\In the present work we examine the zero-order FJ approximation and we leave to future work the discussion of higher order corrections. 
\\

By integrating Eq.~(\ref{eqn:NP}) in the $y$ coordinate and using (\ref{eq:FJhypo}) as a closure  for $c_{\pm}(x,y)$, an effective one-dimensional equation is obtained,

\begin{equation}\label{eq:FJdim}
J_{\pm} =   D c_{\pm}(x)\left[ \frac{\partial_x \int_{-h(x)}^{+h(x)} e^{\mp \beta e \Phi(x,y)}dy}{\int_{-h(x)}^{+h(x)} e^{\mp\beta e \Phi(x,y)} dy} \right] - D \partial_x c_\pm(x)
\end{equation}
where $J_{\pm} = \int dy\ j_{\pm}$ is the longitudinal mass flux per  unit width for each species.
In Eq.~(\ref{eq:FJdim}) the concentrations $c_{\pm}(x)$ are the marginal ones; in the following, we refer to the marginal concentrations unless the both $x-$ and $y-$dependences are explicitly noted.
\\
Now we introduce dimensionless variables. As reported in table \ref{table:rescaledvariables} we rescale the $x$ coordinate by the total length of the channel $L$ and the coordinate $y$ as well as the Debye length $\lambda_D$ and the channel profile $h(x)$ by the half-height $\bar{h}$. In this way the channel profile reads
\begin{equation}
h(x) = 1 + \frac{\kappa}{2} - \kappa x,
\end{equation}
where we also introduced a rescaled channel slope $\kappa= k/\epsilon$. Since we keep fixed the half height $\bar{h}$ allowing for variation in the degree of corrugation we note that $\kappa < 2$ for geometrical consistency. 
\\
The electrostatic potential is rescaled by the thermal one $k_B T/ e$ and the volumetric concentrations $c_{\pm}$ by the concentration in the bulk $c_s$. Consequently, the charge density $q$ is rescaled by $e c_s$, the mass flux $J$ per unit width by $D c_s \bar{h}/L$ and the conductance per unit width, $G=\partial I/\partial \Delta V $ with $I$ the total ionic current and $\Delta V$ the applied potential drop, by the bulk conductance $ D c_s \bar{h}e^2/k_B T L$.
\\
In dimensionless form Eq.~(\ref{eq:FJdim}) now reads:
\begin{equation}\label{eq:FJadim}
J_{\pm} = -  c_{\pm} \Big[ \partial_x \beta A_{\pm} + \partial_x \log c_{\pm} \Big] = - c_{\pm} \partial_x \mu_{\pm}
\end{equation}
where we have introduced the (dimensionless) electrochemical potential $\mu_{\pm} = \log c_{\pm} + \beta A_{\pm}$.
In Eq. (\ref{eq:FJadim}) the electrophoretic contribution now appears in terms of the previously introduced effective free energies $A_{\pm}(x)$ . For a neutral species the effective free energy reduces to the standard Boltzmann entropy $\beta A(x) = -\log 2h(x)$. In this case $\frac{d}{dx}\beta A(x)$ is referred to as an \textit{entropic force}, originating from the variation in phase-space volume available for free diffusion along the channel. For a charged species $A$ embeds both enthalpic and entropic contributions.
\begin{table}[h]
\caption{Adimensionalization of the independent and derived quantities of the model.} \label{table:rescaledvariables}
\begin{tabular}{@{\hskip 0.2in}c@{\hskip 0.2in}c}
\hline\hline 
Variables & Rescaled variables \\ [0.5ex] 
\hline\hline 
Longitudinal coordinate x & x/L  \\ 
Transversal coordinate y & y/$\bar{h}$  \\
Channel profile h(x) & h(x)/$\bar{h}$ \\
Debye length $\lambda_D$ & $\lambda_D $/$\bar{h}$  \\
Channel slope $\kappa$ & k/$\epsilon$ \\  
Electrostatic potential $\phi$ & $e\phi/k_B T$ \\
Volumetric concentrations $c_{\pm}$ & $c_{\pm}$/$c_s$ \\
Charge density $q$ & $q$/$ e c_s$ \\
Mass fluxes $J_{\pm}$ & $J_{\pm}$/$ (D c_s \bar{h}$/$L)$ \\
Differential conductance $G$ & $G$/$ (D c_s \bar{h} e^2$/$ L k_B T)$ \\ [1ex]
\hline 
\end{tabular}
\end{table}
\\
Eq.~(\ref{eq:FJadim}) must be integrated with the appropriate boundary conditions, \textit{i.e.} by imposing continuity in the electrochemical potential at the ends of the channel~\footnote{From Eq.~(\ref{eq:FJadim}) we see that by imposing a continuous electrochemical potential throughout the system we ensure finite fluxes everywhere.}.
\\
The discontinuity in the surface charge distribution at  the channel ends, and the consequent readjustment of ions within the diffusive layer results in an apparent local discontinuity in the concentration and electrostatic potential profiles.  In the present framework, such discontinuities are treated as point-like discontinuities, which stands for the fact that these entrance effects are $\mathit{O}(\epsilon)$, hence they fall within the level of our approximation.
\\
\subsection{Local Debye-H\"{u}ckel approximation}\label{local}
In dimensionless units, the Poisson equation (\ref{eq:Poisson}) reads
\begin{equation}\label{eq:Poissonadim}
\epsilon^2 \partial_x^2 \Phi  + \partial_y^2 \Phi  = - \lambda_D^{-2} q.
\end{equation}
It is  convenient to decompose the electrostatic potential as
\begin{equation}\label{eq:potentialdec}
\Phi(x,y) = \psi(x,y) + \langle \phi \rangle (x)+ \phi_{ext}(x)
\end{equation}
\\
where  $ \langle \phi \rangle = \frac{1}{2h(x)} \int_{-h(x)}^{+h(x)} dy\ \phi(x, y)$ is the average potential across $y$, $\psi = \phi - \langle \phi \rangle$ in the excess potential at each section and $\phi_{ext} = -\Delta V \left( x - \frac{1}{2} \right)$ is the potential drop applied externally, resulting in a constant electric field directed in the $x$ direction.\\
By using FJ approximation into Eq.~(\ref{eq:Poissonadim}) together with (\ref{eq:potentialdec}) and by linearizing in $\psi$ under the assumption of small potential variation in the transversal direction, we reduce (\ref{eq:Poissonadim}) to
\begin{equation}\label{eq:poissonquasifinal}
\epsilon^2 \partial_x^2 \Phi + \partial_y^2 \psi = -	\frac{\lambda_D^{-2}}{2 h} \left[(c_+ - c_-) -(c_+ + c_-)\psi \right].
\end{equation}
\\
We refer to the linearization used to derive Eq. (\ref{eq:poissonquasifinal}) as a \textit{local} Debye-Huckel (DH) approximation: the potential is linearized with respect to the local cross-sectional average preserving therefore global nonlinearity. We stress that the assumption of small $\psi$ is more general than standard DH, which requires small $\zeta$ potential everywhere (typically  \cite{ANDELMAN1995603} $\zeta \le 25 mV$). 
In fact it allows to explore  the \textit{ideal gas} \footnote{the name refers to the fact that in this regime the only contribution to the electrostatic pressure between the two walls come from the entropy of an homogeneous solution of noninteractive ions $ P \sim \frac{1}{h}$} regime for arbitrarily high Dukhin number, where typically the global Debye-H{\"u}ckel assumption would fail.
\\
 The small aspect ratio constraint allows for a lubrication-like approximation of (\ref{eq:poissonquasifinal})  which reduces to a linear equation for  $\psi$:
 \begin{equation}\label{eq:poissonfinal}
 \partial_y^2 \psi = -	\frac{\lambda_D^{-2}}{2 h} \left[(c_+ - c_-) -(c_+ + c_-)\psi \right]
\end{equation}
 \\
Consistently, the scaling argument applies as well to the electrostatic wall boundary condition, which after neglecting terms of $\mathcal{O}(k^2)$, and introducing rescaled variables, reduces to
\begin{equation}\label{eq:electrbc2}
\partial_y \phi \Big|_{y = \pm h} = \mp \frac{Du\ }{\lambda_D^2}.
\end{equation}
We shall note here that the LTE hypothesis previously introduced implies local electroneutrality, in which the integrated charge density balances the surface charge density at each cross-section,
\begin{equation}\label{eq:electroneutral}
\int dy\ q(x,y) = c_+(x) - c_-(x) = 2\ Du.
\end{equation}
In fact, by integrating Eq.~(\ref{eq:poissonquasifinal}) in the $y-$coordinate, using (\ref{eq:electrbc2})
and neglecting $\mathcal{O}(\epsilon^2)$ terms, Eq. (\ref{eq:electroneutral}) is obtained.
\\
 To first order approximation the FJ ansatz, the lubrication approximation and  local electroneutrality are different naming for the same unique assumption, \textit{i.e.} separation of transversal and longitudinal scales, applied to different physical properties\cite{doi:10.1063/1.1669549}.
The reduced Poisson equation can be formally integrated leading to
\begin{equation}\label{eq:potential}
\psi= -\frac{Du(x)}{\lambda_D(x)} \frac{\cosh(y/\lambda_D(x))}{\sinh(h(x)/\lambda_D(x))}+ \frac{c_+(x) - c_-(x)}{c_+(x) + c_-(x)} .
\end{equation}

In Eq.~(\ref{eq:potential}) the potential is naturally expressed in terms of a local Dukhin number and a local Debye length respectively defined as:
\begin{eqnarray}
\lambda_D(x) = \frac{\lambda_D}{\sqrt{c_{vol}(x)}} \label{eq:localDeb}\\
Du(x) = \frac{Du}{c_{vol}(x)} \label{eq:localDu}
\end{eqnarray}
in terms of the total average volumetric concentration $c_{vol}(x) = [c_+(x) + c_-(x)]/2 h(x)$.\\
We recognize the first term on the \textit{rhs} of Eq.~(\ref{eq:potential}) to be the Debye-Huckel potential carrying an extra $x-$dependence due to the varying channel geometry. The second term on the \textit{rhs} ensures local electroneutrality.
\\ 
Eqs.~(\ref{eq:FJadim}) and (\ref{eq:poissonfinal}) need to be solved numerically.
It is convenient to rewrite Eq.~(\ref{eq:FJadim}) in terms of $\psi$ :
\begin{widetext}
\begin{subequations}
\begin{eqnarray}\label{eq:systemcomsol}
J_+ &=& -\partial_x c_+ + c_+ [\partial_x \log h -(\partial_x \langle \phi \rangle - \Delta V ) +\partial_x \log \langle e^{-\psi} \rangle ] \label{subeq:plus}\\
J_- &=& -\partial_x c_- + c_- [\partial_x \log h +(\partial_x \langle \phi \rangle - \Delta V ) +\partial_x \log \langle e^{+\psi} \rangle ] \label{subeq:minus} \\
\partial_y^2 \psi &=& -\frac{\lambda_D^{-2}}{2 h} [(c_+ - c_-) - (c_+ + c_-) \psi]  \label{subeq:psi}
\end{eqnarray}
\end{subequations}
\end{widetext}
so that the coupling between the concentration profiles and the electrostatic potential is made now explicit.
We use finite-element simulations (COMSOL) to solve the system of Eqs.~(\ref{subeq:plus})(\ref{subeq:minus})(\ref{subeq:psi}) in order to look at the electric current $I = J_+ - J_-$ generated by the applied potential drop $\Delta V$. (See appendix for details on the numerical simulations).
\\
 The expression for the electric current obtained by formally integrating Eq.~(\ref{subeq:plus}) and (\ref{subeq:minus}) reads
\begin{equation}\label{eq:current}
I = -\frac{1}{2} \left[\frac{e^{-\Delta V/2}- e^{+\Delta V/2}}{\displaystyle\int_{0}^{1} dx'\ e^{\beta A_+(x')}} - \frac{e^{+\Delta V/2}- e^{-\Delta V/2}}{\displaystyle\int_{0}^{1} dx'\ e^{\beta A_-(x')}} \right],
\end{equation}
where the denominator is responsible for the non-linear (rectified) response of the channel, as  it expresses the coupling between the dissipative dynamics (thermodynamic forcing) and the geometric asymmetry. For a flat channel, Eq.~(\ref{eq:current}) reduces to the standard ohmic response\cite{RevModPhys.80.839} (per unit width)  which in dimensional unit reads
\begin{eqnarray}
I_{ohm}  = \frac{D e^2}{k_B T} \left[\frac{2 c_-}{L} + \frac{2 \sigma}{e L}\right]\ \Delta V.
\end{eqnarray} 
\\
Eq.~(\ref{eq:current}) is valid for slowly varying channels under the assumption of small potential variation in the transversal direction. Hence it represents a well-grounded expression for the ionic current allowing to span across different regimes in the electrostatic phase space in both $\lambda_D$ and $Du$. 
\\ Previously proposed analytical approaches~\cite{B301021J, PhysRevE.77.031131,  doi:10.1063/1.2179797}   assume that $\lambda_D$ is  the relevant controlling parameter by treating separately the case of no overlap $\lambda_D \ll 1$ and strong overlap $\lambda_D \gg 1$. This is not necessary in the present framework, where   $\lambda_D$ can vary continuously. Nevertheless, it is useful at this stage to introduce the regime of strong Debye overlap as it represents a well-known scenario which we will use as a benchmark to compare with  numerical results. 
\subsection{Strong Debye overlap, $\lambda_D \gg 1$}
Let us  consider the regime in which the channel height is much smaller than the Debye length. The EDL extends all throughout the interior of the confined electrolyte, rendering the channel perfectly charge-selective. Both the electrostatic potential, $\Phi(x)$, and the ionic concentration profiles, $c_{\pm}(x)$, are assumed to be uniform in the transversal direction allowing for a substantial simplification of the mathematical problem at hand. We stress that the concentrations $c_{\pm}(x)$ here are not the marginal  concentrations but the total concentrations which in this limit are independent of $y$.
\\
Together with local electroneutrality which in this case reads
\begin{equation}\label{eq:LEDonnan}
2h(x) \left[ c_+(x) - c_-(x) \right] = 2 Du,
\end{equation}
 continuity in the chemical potential provides an expression for the Donnan potential at either end of the channel~\cite{PhysRevLett.111.244501, RevModPhys.80.839, PhysRevE.76.041202},
\begin{subequations} \label{eq:Donnan}
\begin{eqnarray}
\phi_{L} = \frac{1}{2} \log \left[ \frac{-Du + \sqrt{Du^2 + h_L^2}}{+ Du + \sqrt{Du^2 + h_L^2}} \right] + \frac{\Delta V}{2}, \\
\phi_{R} = \frac{1}{2} \log \left[ \frac{-Du + \sqrt{Du^2 + h_R^2}}{+ Du + \sqrt{Du^2 + h_R^2}} \right] - \frac{\Delta V}{2}.
\end{eqnarray}
\end{subequations}
Notably, already at equilibrium the varying geometry results in a non-uniform \textit{tilted} potential across the channel. 
\\
Analogously, the channel's junctions concentrations read
 \begin{subequations}
\begin{eqnarray}\label{eq:systemDonnan}
c_L = c_L^+ + c_L^- =\frac{ \sqrt{Du^2 + h_L^2 }}{h_L}, \\\label{subeq:leftdonnan}
c_R =  c_R^+ + c_R^- = \frac{ \sqrt{Du^2 + h_R^2}}{h_R}. \label{subeq:rightdonnan}
\end{eqnarray}
\end{subequations}
Hence,  a jump in concentration profiles builds up at each junction of the channel to compensate for the potential discontinuity (\ref{eq:Donnan}). Such a local balance is known in the literature  as \textit{local} Donnan equilibrium. These expressions will allow for  asymptotic analytical predictions for the  conductances when  $\Delta V \to \pm \infty$.
\\
The equation of motion (\ref{eq:FJadim}) for $\lambda_D \gg 1$ reduces to
\begin{equation}
2 h(x) \left[ \mp c_{\pm}(x) d_x \phi(x) -  d_x c_{\pm}(x) \right]= J_{\pm},
\end{equation}
which, rewritten in terms of the total mass flux $J= J_+ + J_-$ and electric current $I$, becomes
\begin{subequations}
\begin{eqnarray}\label{subeq:NPstrongoverlap1}
J &=& - 2 h(x) d_x c(x) - 2 Du\ d_x \phi, \\
\label{subeq:NPstrongoverlap2}
I &=& - 2 h(x) c(x) d_x \phi + 2 Du\ d_x \log 2 h(x).
\end{eqnarray}
\end{subequations}
In (\ref{subeq:NPstrongoverlap1}-\ref{subeq:NPstrongoverlap2}) local electroneutrality (\ref{eq:LEDonnan})  has been used to further simplify the expressions.
\section{RESULTS}
\subsection{Current response and limiting conductances}
\begin{figure}[h]
\includegraphics[width=9cm]{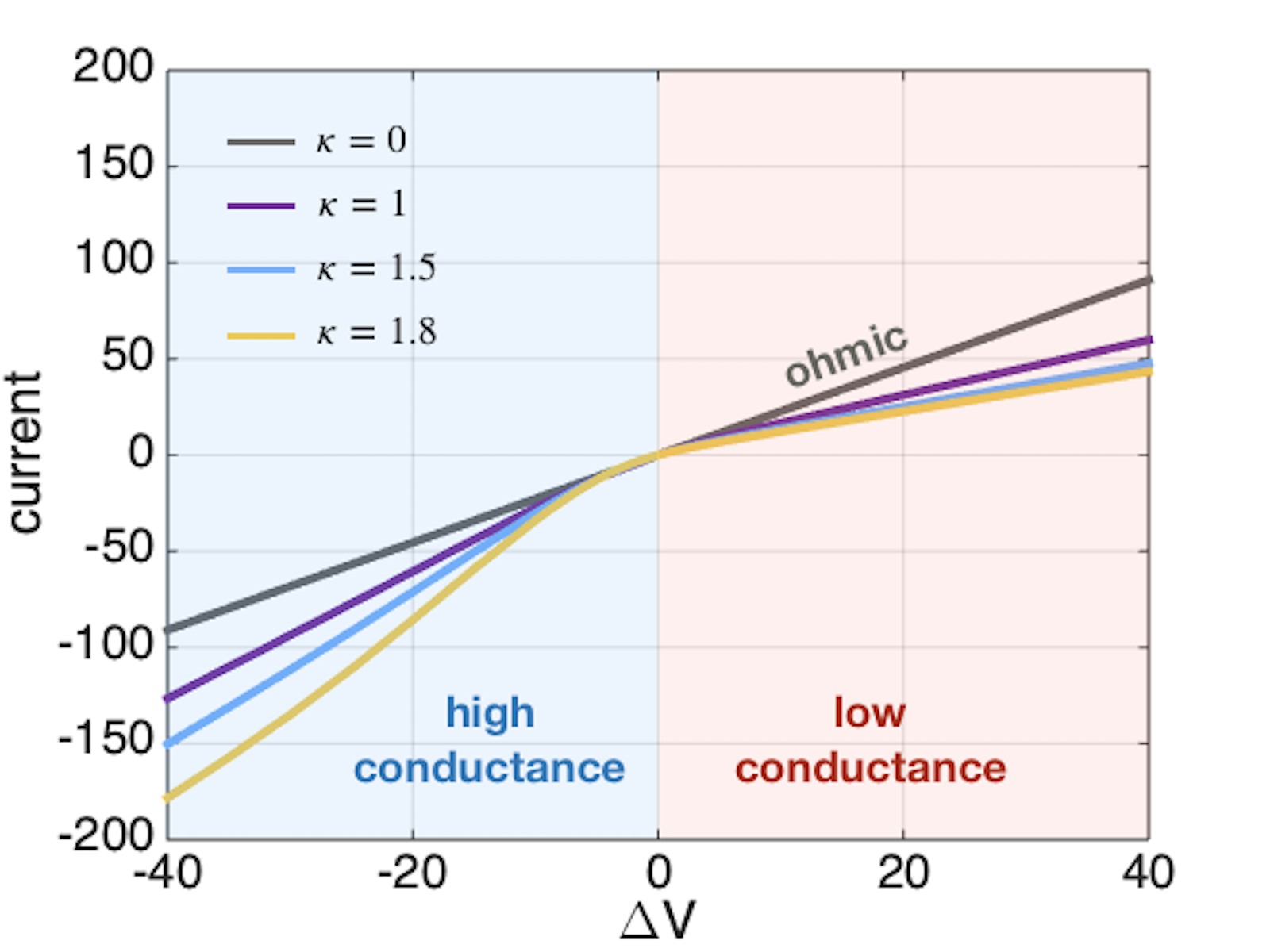}
\caption{\label{FIG:IV_Debye0p5} Dimensionless current I as a function of $\Delta V$ for a channel with $\lambda_D = 1/2$, $Du =1/2$ at different value of channel slope, respectively $\kappa=0, 1, 3/2, 1.8$. We recognize two different conductance states. For positive voltage drop (positive electric field) the system is in a low conductance state, the current being smaller than the Ohmic one (grey line). On the contrary for negative voltage drop (negative electric field) the current is magnified and the system is said to be in a high conductance state.
}
\end{figure}

\begin{figure}[h]

\includegraphics[width=9cm]{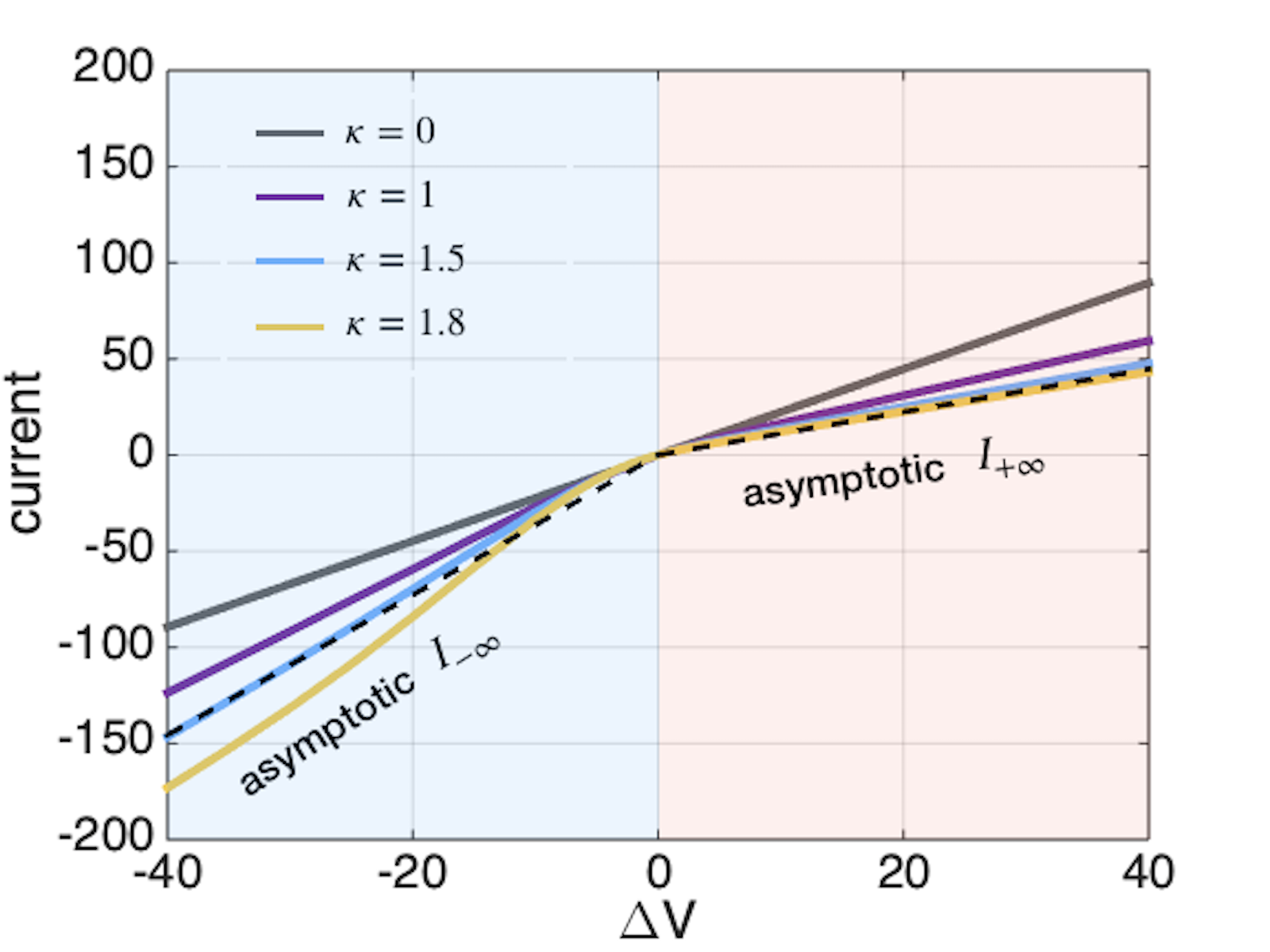}
\caption{\label{FIG:IV_Debye2} Dimensionless current $I$ as a function of $\Delta V$ for a channel with $\lambda_D = 2$, $Du =1/2$ at different value of channel slope, respectively $\kappa=0, 1, 1.5, 1.8$. The black dashed lines are the limiting currents for $k =3/2$ in the limit of $\Delta V \to \pm\infty$ using Eqs.~(\ref{eq:limitingGExplicit}). 
}
\end{figure}

We focus first on the current response obtained by numerically solving the system (\ref{eq:systemcomsol}-\ref{subeq:psi}) under an applied potential difference $\Delta V$. The two reservoirs are kept at the same ionic strength so that the only thermodynamic force at play is a constant electric field along the longitudinal coordinate. A positive (negative) $\Delta V$ corresponds respectively to the anode being placed at the left (right) reservoir.\\
A standard measure of ionic rectification is given by the current-voltage (I-V) curve which we report in Fig.~(\ref{FIG:IV_Debye0p5})  for the case of $\lambda_D = 1/2$ and $Du = 1/2$ and for different values of the channel slope. This regime corresponds to the case of partial Debye overlap inside the channel. For instance, in the case of $k=3/2$ the local ratio $\frac{\lambda_D}{h(x)}$ spans from $\sim 0.3$ at the base junction up to $\sim 2$ at the tip junction. Therefore by moving from left to right ions experience the building up of a Donnan potential  passing from a region (left) where the bulk dominates to a region (right) where the EDL dominates.
The non-linear curves in Fig.~(\ref{FIG:IV_Debye0p5}) display the usual diode-like behaviour reported in the literature,  with a preferential direction of ionic current. When the electric field is applied parallel to the $x-$direction with the counterions moving from base to tip, the current is suppressed with respect to the Ohmic response (grey curve) and the system is said to be in a  \textit{low conductance state}. On the contrary, when the electric field is applied antiparallel with respect to $x$ with the counterions moving from tip to base, the current is magnified and the system is said to be in a \textit{high conductance state}. 
\\
The  rectification magnitude is monotonous in the degree of asymmetry in the system. The greater the channel's slope the larger the rectification. This must come as no surprise since the channel slope is the only element introducing asymmetry in the system. For $k \to 0$ the channel is flat and it behaves like a standard Ohmic resistor.
\\
The numerical I-V curves can be compared with analytical predictions of the limiting differential conductances
\begin{equation}
G_{\pm \infty} = \lim_{\Delta V \to \pm \infty} \frac{\partial I}{\partial \Delta V}.
\end{equation}
For strong Debye overlap  the equations of motion  reduce to (\ref{subeq:NPstrongoverlap1}-\ref{subeq:NPstrongoverlap2}). By neglecting the diffusive contribution to the mass flow with respect to the electrophoretic contribution in (\ref{subeq:NPstrongoverlap1}) and by integrating in $x$ we obtain
\begin{equation}
J = 2 Du \Delta V.
\end{equation}
Combining Eqs.~(\ref{subeq:NPstrongoverlap1}) and (\ref{subeq:NPstrongoverlap2}) we solve for $d_x c$ in terms of the ratio $\frac{I}{J}$
\begin{equation}\label{eq:expressionJI}
2 h d_x c + \frac{(2 Du)^2 }{2 h c} d_x\log 2h = \left( \frac{2 Du}{2h c} \frac{I}{J}-1 \right)J,
\end{equation}
which is bound asymptotically,  $\Delta V \to \pm \infty$, if the prefactor on the \textit{rhs} vanishes, \textit{i.e.} $\frac{2 Du}{2h c} \frac{I}{J} \to 1$. Accordingly  the limiting conductance, $G_{\pm \infty}$, reduces to
\begin{equation}\label{eq:limitingG}
G_{\pm \infty} = \lim_{\Delta V \to \infty}(c_+ + c_-),
\end{equation}
because  the diffusive contribution to the ionic flux for very large fields is negligible. Eq.~(\ref{eq:limitingG}) implies that the marginal concentration inside the channel approaches a uniform value in the limit $\Delta V \to \pm \infty$.
When $\lambda_D \gg 1$ we have analytical expressions for the marginal concentration at the channel's ends where, due to the channel geometry, the left end is characterized by the higher marginal concentration while the right end fixes the lower value. Hence,  from Eqs.~(\ref{subeq:leftdonnan}-\ref{subeq:rightdonnan}) (see Discussion section for further details)
\begin{subequations}\label{eq:limitingGExplicit}
\begin{eqnarray}
G_{+\infty} = c_{vol}^{R} = 2 \sqrt{Du^2 + h_R^2}\label{subeq:limitingGexpliticL} ,\\
G_{-\infty} = c_{vol}^{L} = 2 \sqrt{Du^2 + h_L^2}\label{subeq:limitingGexpliticR}.
\end{eqnarray}
\end{subequations}
In Fig.~(\ref{FIG:IV_Debye2}) we show the I-V curves for $\lambda_D = 2 $ and $ Du =1$, \textit{i.e.} in the regime of strong overlap. For $k = 3/2$ we report the analytical predictions for the asymptotic curves $I_{\pm \infty} =\pm G_{\pm \infty} \Delta V $ 
with the limiting conductances obtained from (\ref{eq:limitingGExplicit}), showing that these  analytical expressions accurately capture the numerical results. Further discussion on the saturation mechanism for the conductance are reported in the Discussion section.
\\
From the  comparison between Fig.~(\ref{FIG:IV_Debye0p5}) and Fig.~(\ref{FIG:IV_Debye2}) we observe that the quantitative structure of the I-V curves does not change respectively for partial Debye overlap with $\lambda_D =1/2$ and strong overlap with $\lambda_D =2$.  It follows that the Debye length seems not to play a primary role in governing rectification. Notably, this is at odd with previous understanding of ICR which relies on $\lambda_D$ as the main controlling parameter. 
\\In the next session this observation is further explored and clarified by looking closely to the dependence of ICR on the electrostatic lengthscales.
\\

\subsection{Current rectification ratio}
\begin{figure}
    \centering
    \includegraphics[width=9 cm]{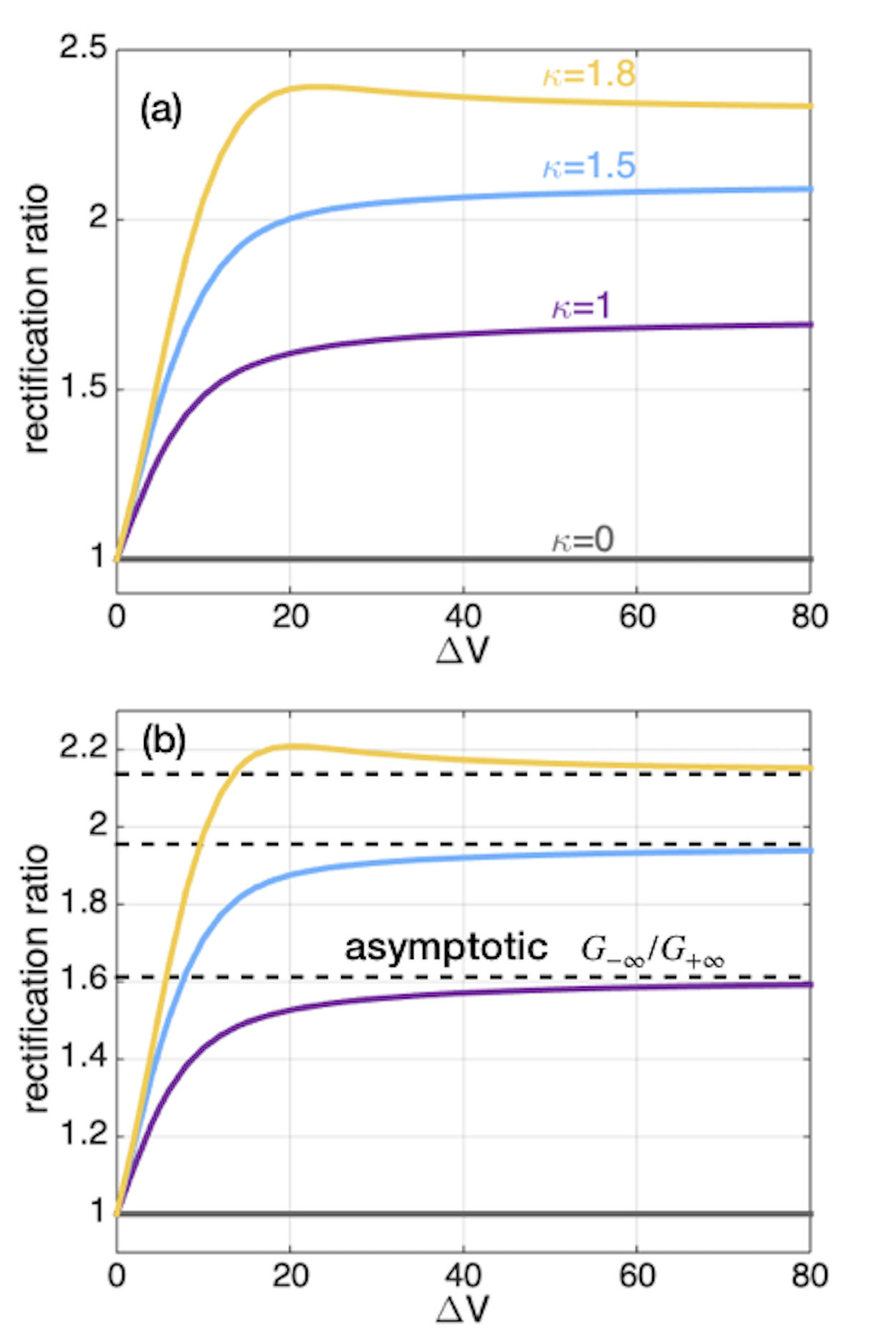}
    \caption{(a) The rectification ratio $\eta$ as a function of the amplitude $|\Delta V|$ in the case of $\lambda_D =1/2$ and $Du =1$ for $\kappa = 0$, $\kappa = 1$, $\kappa =3/2$ and $\kappa = 1.8$. For a flat channel $\eta =1$ and the response is linear (gray line). (b) The rectification ratio $\eta$ as a function of the amplitude $|\Delta V|$ in the case of $\lambda_D =2$ and $Du =1$ for $\kappa = 0$, $\kappa = 1$, $\kappa =3/2$ and $\kappa = 1.8$. The dashed black lines show the asymptotic value for $\eta$ in the limit of $|\Delta V| \to \infty$. }
    \label{FIG:etaVSDV}
\end{figure}

In order to gain further insights on the rectified behaviour of the present system we introduce the rectification ratio $\eta$
\begin{equation}
\eta = \frac{|I(-\Delta V)|}{|I(+\Delta V)|},
\end{equation}
defined as the ratio between the absolute value of the current for opposite polarity of the external field. In the case of an ohmic resistor $\eta =1$.
\\
Fig.~(\ref{FIG:etaVSDV}) displays $\eta$ as a function of the external forcing, $\Delta V$, for $\lambda_D =1/2$ and $\lambda_D =2$. Each plot shows the rectification ratio for different value of the channel slope. The asymptotic predictions for $\eta$ obtained from Eqs.~(\ref{subeq:limitingGexpliticL}) and (\ref{subeq:limitingGexpliticR}) are reported in Fig.(\ref{FIG:etaVSDV}-b) (dashed black lines). 
\\

\begin{figure}
\includegraphics[width=9cm]{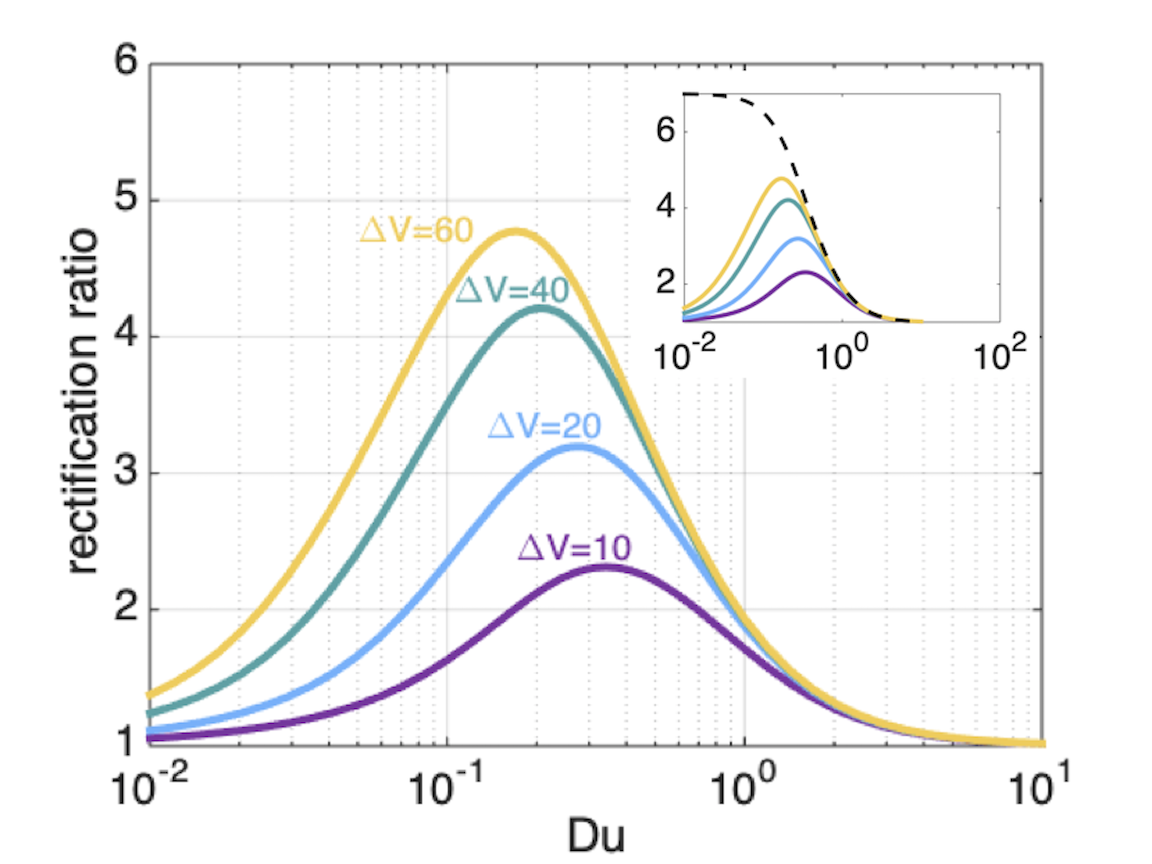}
\caption{\label{FIG:etaVSDu} The rectification ratio $\eta$ as a function of the reference $Du$ for a channel with $\lambda_D = 2$ and $k =3/2$ at different value of external forcing, respectively $\Delta V=10, 20, 40, 60$. In the inset graph the analytical prediction for $\eta$ in the regime of strong overlap and of $|\Delta V| \to \infty$ is reported. For sufficiently large $Du$ it accurately estimates the behaviour of $\eta$ while in the limit of $Du \to 0$ it deviates from the numerical curves because of the breakdown of the hypothesis of strong overlap. 
}
\end{figure}

\begin{figure}
\includegraphics[width=9cm]{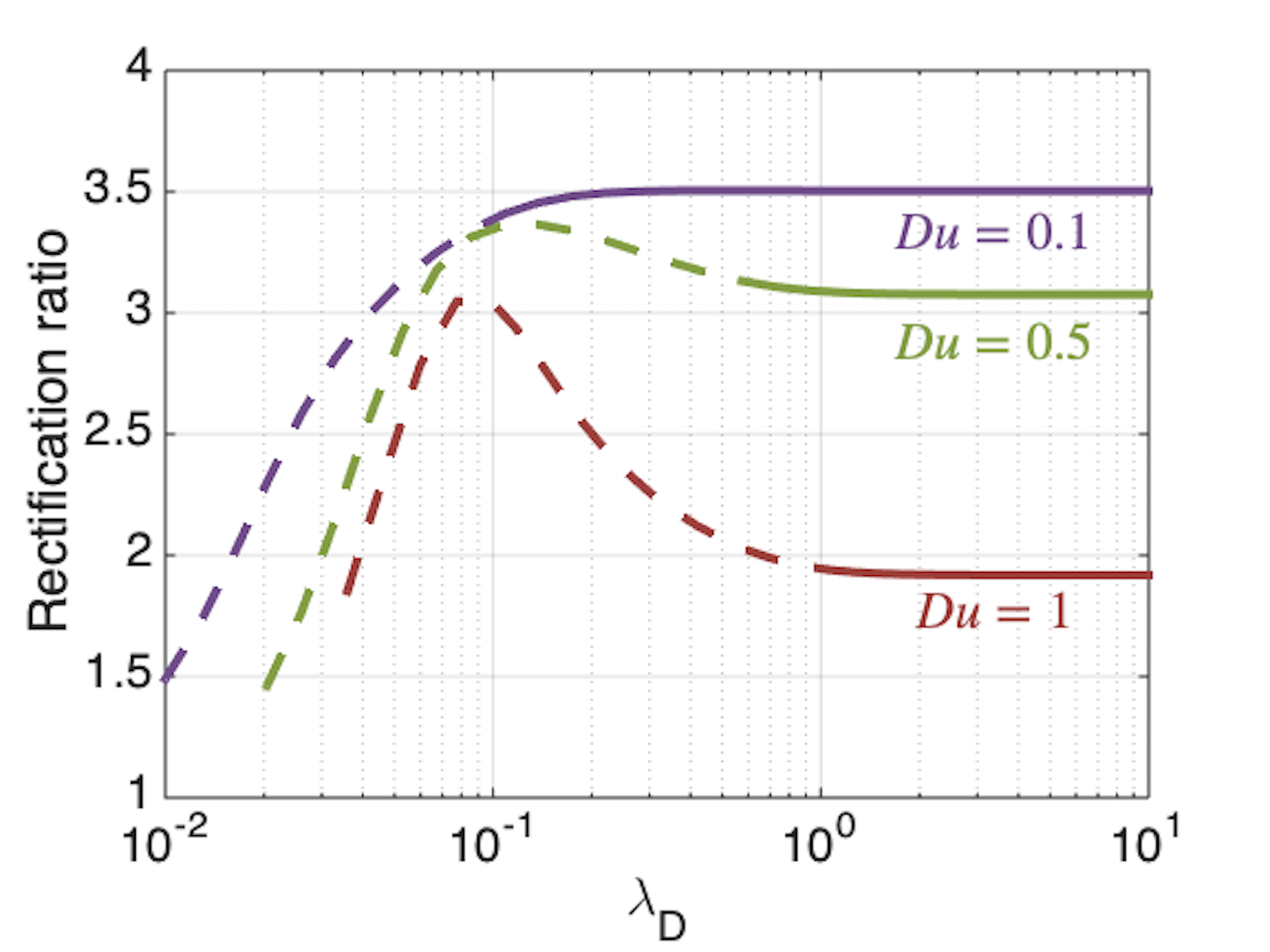}
\caption{\label{FIG:etaVSlambdaD} The rectification ratio $\eta$ as a function of the Debye length for different value of the Dukhin, respectively $Du = 1/10$, $Du=1/2$, $Du=1$. The channel slope is $\kappa = 3/2$ and the potential drop $\Delta V = 40$. Dashed lines refer to the regime in which the approximation of local Debye-H{\"u}ckel approximation is no longer justified. 
}

\end{figure}

Fig.~(\ref{FIG:etaVSDV}) shows a saturation behaviour for large value of $\Delta V$. The saturation value increases with the channel slope as already observed for the I-V curves. Moreover for strong overlap the analytical expressions (dashed lines) are in good agreement with the numerical results. 
\\
We now turn our attention to the dependence of ICR on the Dukhin number.
Fig~(\ref{FIG:etaVSDu})  shows $\eta$ as a function of the reference Dukhin number, Eq.~(\ref{eq:refDu}), for $\lambda_D = 2$ and $\kappa = 3/2$ for different values of the external forcing. Interestingly, $\eta$ shows a strongly non-monotonic dependence on $Du$ with a maximum of rectification approximately at $Du \approx 1/2$. For $Du \ll 1$ or $Du \gg 1$ the rectification ratio goes to one and the standard ohmic behaviour is recovered. For values of $Du$ close to unity the rectification ratio reaches a maximum which depends on the strength of the applied field upon reaching a saturation value as shown in Fig.~(\ref{FIG:etaVSDV}). The saturation value of $\Delta V$ is itself modulated by $Du$.
Fig.~(\ref{FIG:etaVSDu}) shows that $Du$ is a critical parameter controlling rectification, in contrast with $\lambda_D$ that seems not to be an adequate parameter to describe ICR. This is further illustrated by looking at Fig.~(\ref{FIG:etaVSlambdaD}), where $\eta$ is plotted as a function of $\lambda_D$ for three different values of $Du$. We report a dashed line when we enter the regime in which linearization in $\psi$ is no further justified. This happens in the limit of small $\lambda_D$ when the potential at the centerline vanishes and $\psi \sim \zeta$. In the regime of partial and strong overlap no significant dependence on $\lambda_D$ is shown. Albeit not quantitative, our results suggest that ICR decreases while approaching the limit of vanishing $\lambda_D$. In this limit it is known that ICR approaches a non-zero asymptotic value\cite{doi:10.1021/acs.jpcb.8b11202}. 
\\
\section{DISCUSSION: The role of Dukhin number}
IThe results of the previous section show that ICR is not primarily governed by the Debye length but rather by the Dunkhin length. This suggests that the Dukhin number directly controls the high (low) conductance state,  for negative (positive) potential drop.
\begin{figure}
    \centering
    \includegraphics[width=9 cm]{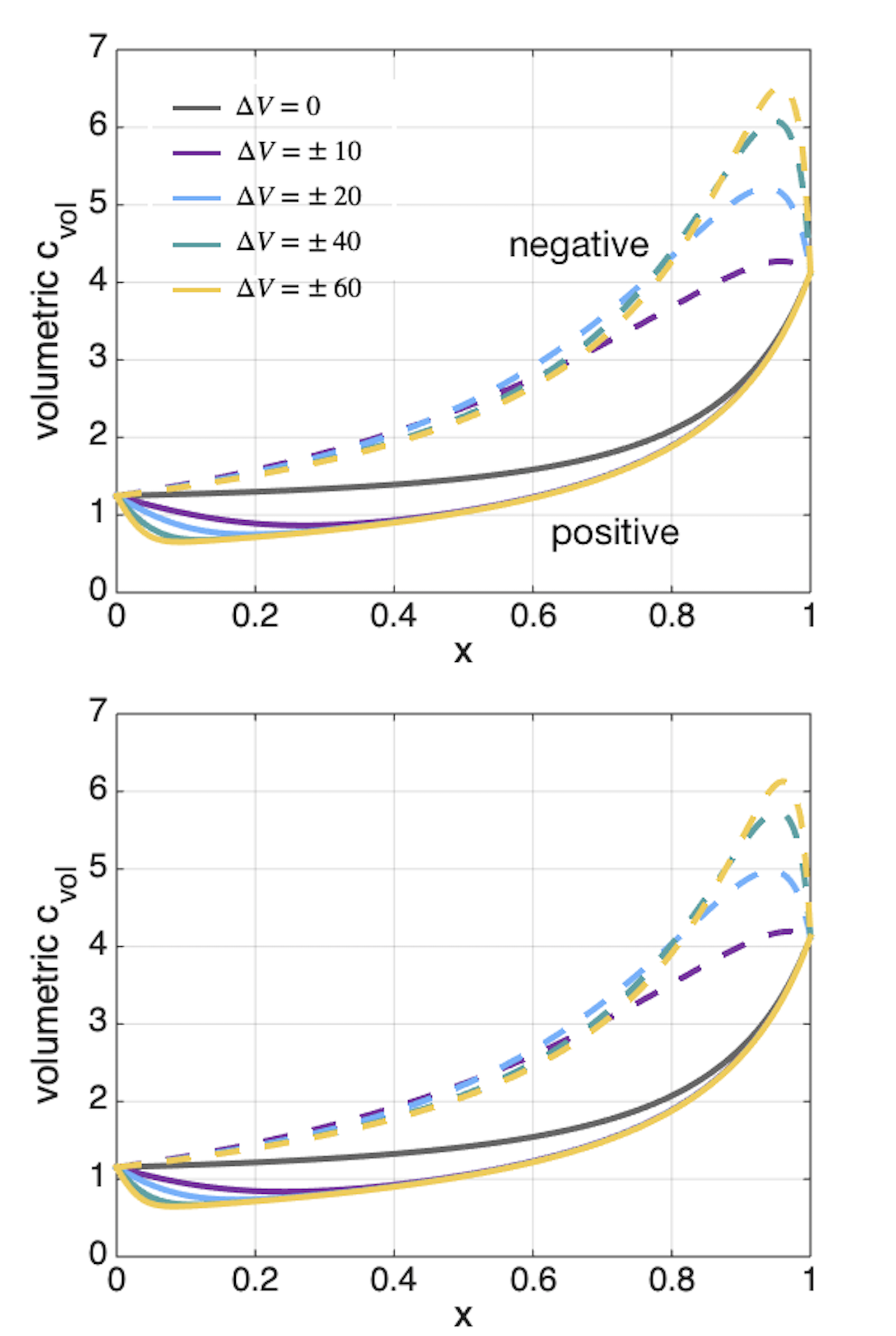}
    \caption{ Volumetric cross-section average concentration $c_{vol}$ along the channel axis for different amplitude of the applied potential, respectively $\Delta V = 0$, $\Delta V = |10| $, $\Delta V = |20| $, $\Delta V = |40| $ and $\Delta V = |60| $. In the figures solid lines corresponds to a positive potential drop while dashed lines to a negative potential drop. Concentration profiles for  the following choice of parameters:  (a)  $\lambda_D = 1/2$, $Du = 1$ and $\kappa = 3/2$  (b) $\lambda_D = 2$, $Du = 1$ and $\kappa = 3/2$. }
    \label{FIG:c_vol}
\end{figure}
\begin{figure*}
    \centering
    \includegraphics[width=17 cm]{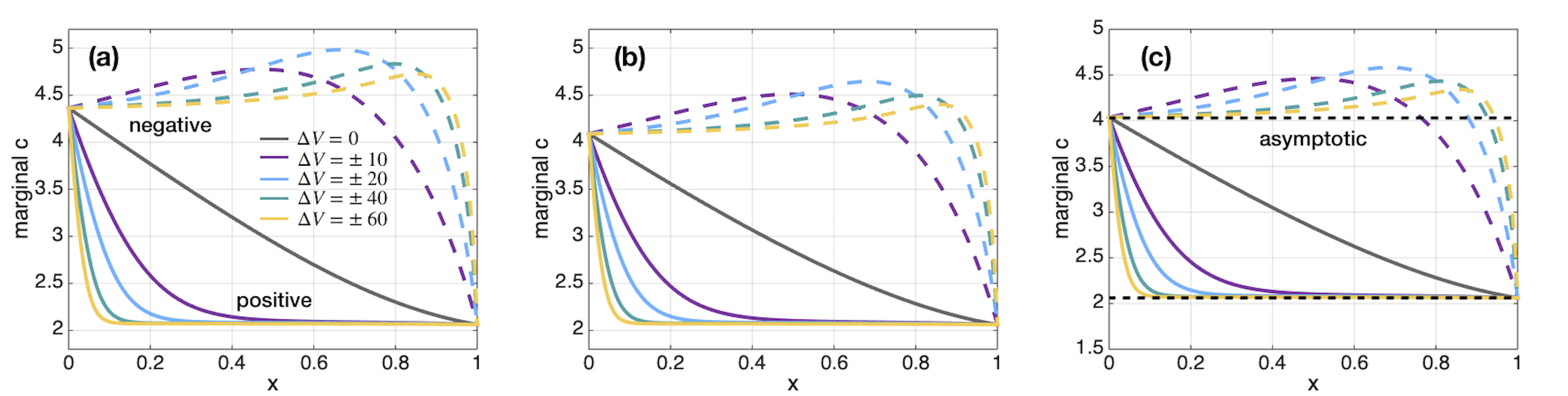}
    \caption{ Marginal concentration $c$ along the channel axis for different amplitude of the applied potential, respectively $\Delta V = 0$, $\Delta V = |10| $, $\Delta V = |20| $, $\Delta V = |40| $ and $\Delta V = |60| $. In the figures solid lines corresponds to a positive potential drop while dashed lines to a negative potential drop. Concentration profiles for  the following choice of parameters: (a) $\lambda_D = 1/2$, $Du = 1$ and $\kappa = 3/2$. (b) $\lambda_D = 1$, $Du = 1$ and $\kappa = 3/2$ (c) $\lambda_D = 2$, $Du = 1$ and $\kappa = 3/2$. The black dashed lines in (c) corresponds to the boundary value for the marginal concentration due to the local Donnan equilibrium (\ref{subeq:limitingGexpliticL}) and (\ref{subeq:limitingGexpliticR}). }
    \label{FIG:c_marg_deb}
\end{figure*}
\begin{figure*}
    \centering
    \includegraphics[width=17 cm]{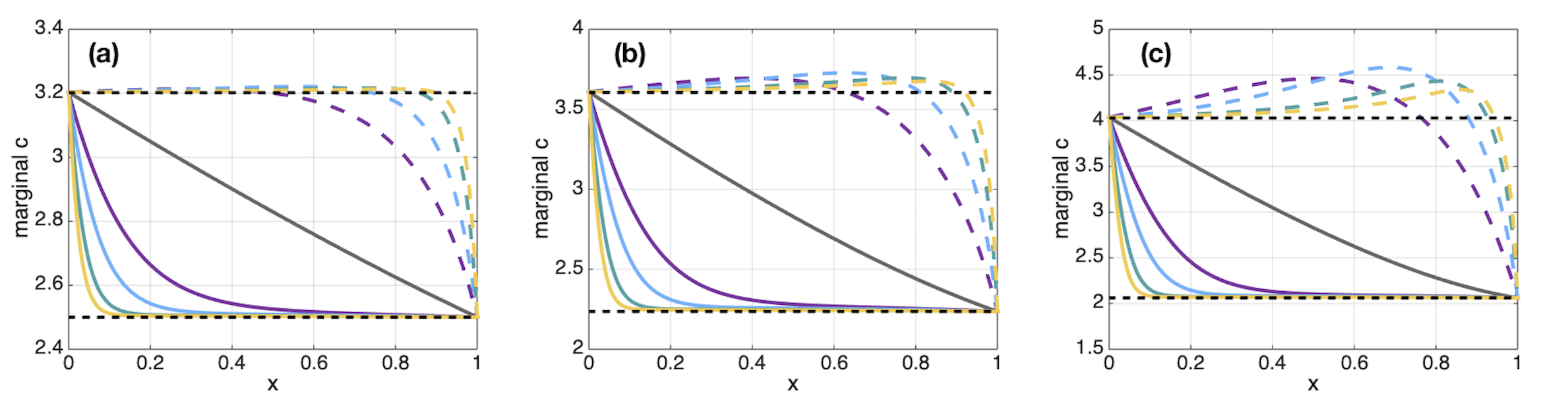}
    \caption{ Marginal concentration $c$ along the channel axis for different amplitude of the applied potential, respectively $\Delta V = 0$, $\Delta V = |10| $, $\Delta V = |20| $, $\Delta V = |40| $ and $\Delta V = |60| $. In the figures solid lines corresponds to a positive potential drop while dashed lines to a negative potential drop. Concentration profiles for  the following choice of parameters: (a) $\lambda_D = 2$, $Du = 1$ and $\kappa = 1/2$. (b) $\lambda_D = 2$, $Du = 1$ and $\kappa = 1$ (c) $\lambda_D = 2$, $Du = 1$ and $\kappa = 3/2$. The black dashed lines corresponds to the boundary value for the marginal concentration due to the local Donnan equilibrium (\ref{subeq:limitingGexpliticL}) and (\ref{subeq:limitingGexpliticR}). }
    \label{FIG:c_marg}
\end{figure*}
This can be understood in terms of ionic concentration enrichment and depletion for opposite polarity of the external field, as discussed in previous works \cite{B301021J, doi:10.1002/adfm.200500471, doi:10.1063/1.2179797}. The panel in Fig.~(\ref{FIG:c_vol}) shows the volumetric cross-sectionally averaged concentration $c_{vol}$ along the channel axis for two different regimes of $\lambda_D$ . In both figures we observe an overall increase (decrease) of ionic concentration for negative (positive) $\Delta V$ with respect to the equilibrium profile, represented by the grey line. Therefore the high conductance state for negative $\Delta V$ is due to an increase in ionic concentration inside the channel. The larger the external forcing, the stronger the accumulation of ions. On the contrary, when a positive voltage drop is applied the electrical conductance decreases due to the decrease of ionic concentration.
\\
In order to understand the phenomenon of salt accumulation and depletion  we now turn our attention to the behaviour of the marginal concentration for large fields. In the previous section we already anticipated that in the limit of very large potential drop we expect the marginal concentration to saturate to a uniform value along the channel axis. Fig.~(\ref{FIG:c_marg_deb}) reports the marginal concentration along the longitudinal axis for increasing value of $\lambda_D$ ((a)-(c)). For increasing amplitude of the external forcing the marginal concentration indeed tends to a constant value which is determined by the boundary value at either end of the channel. In the case of a negative potential drop the marginal concentration saturates to the larger boundary value which is the value at the left end of the channel (base). On the other hand, for positive potential drop the saturation value is bounded to the boundary condition at the right site (tip). 
\\
Fig.~(\ref{FIG:c_marg_deb}) also shows an overshoot in the marginal concentration for large (but finite) negative $\Delta V$. The overshoot is not present in the case of positive $\Delta V$ which stands as an additional sign of the asymmetry in the system. The microscopic mechanism causing it is still not clear and requires further investigations. Fig.~(\ref{FIG:c_marg}) reports the marginal concentration profiles for increasing slope of the channel showing a significant dependence of the overshot on $\kappa$.
\\
 
\begin{figure}[h!]
    \centering
    \includegraphics[width=9 cm]{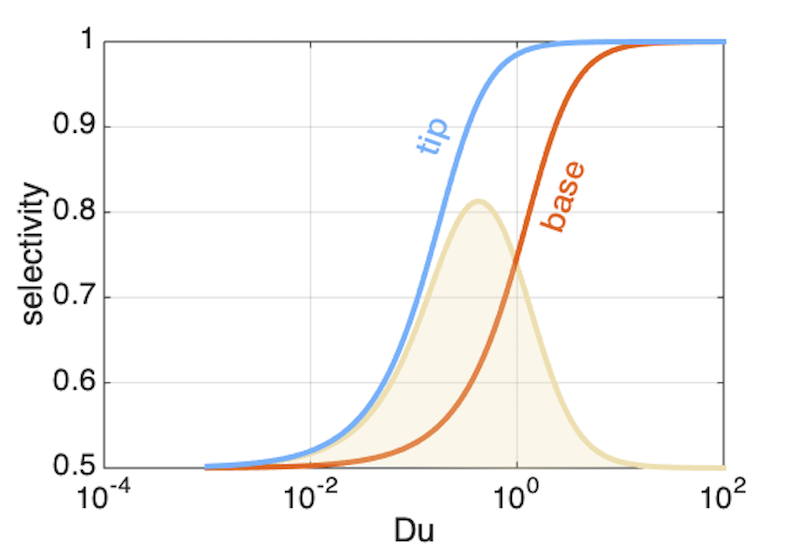}
    \caption{The counterion selectivity at the tip $\gamma_+^R$(blu) and at the base $\gamma_+^L$(red) of a channel with $k=3/2$ as a function of the reference Dukhin number. For $Du \sim 1/2$ the difference between the two selectivities (yellow curve) is maximized, leading to a maximum of rectification.  }
    \label{FIG:selectiv}
\end{figure}

Fig.~(\ref{FIG:c_marg_deb})(c) displays the marginal concentrations for strong overlap, $\lambda_D \gg 1$. Local Donnan equilibrium builds up  at the  nanopore ends, controlling the  corresponding marginal concentrations 
\begin{subequations}\label{eq:marginalDonnan}
\begin{eqnarray}
c_L = 2 h_L \sqrt{\left(\frac{Du}{h_L}\right)^2 + 1}, \\
c_R = 2 h_R \sqrt{\left(\frac{Du}{h_R}\right)^2 + 1}. 
\end{eqnarray}
\end{subequations}
Asymptotically,  $Du \gg 1$, $c_L \to c_R$, \textit{i.e.} $\eta \to 1$. In this regime transport is controlled by the diode surface,  where entropic interactions are negligible with respect to electrostatic interactions and ions do not feel the symmetry breaking originated from the confinement. That is to say, enthalpy wins. 
\\
The local marginal selectivity, $\gamma_{\pm}(x)$ (directly proportional to the ionic  marginal concentrations), constitutes a second, relevant  quantity. For the counterions, the local selectivity at either end of the channel respectively reads
\begin{subequations}\label{eq:selectivities}
\begin{eqnarray}
\gamma_+^{L} = \frac{c_+^L}{(c_+^L + c_-^L)} = \frac{\frac{Du}{h_L}  + \sqrt{\left(\frac{Du}{h_L}\right)^2 + 1}}{2 \sqrt{\left(\frac{Du}{h_L}\right)^2 + 1}} \\
\gamma_+^{R} = \frac{c_+^R}{(c_+^R + c_-^R)} = \frac{\frac{Du}{h_R}  + \sqrt{\left(\frac{Du}{h_R}\right)^2 + 1}}{2 \sqrt{\left(\frac{Du}{h_R}\right)^2 + 1}}
\end{eqnarray}
\end{subequations}
making transparent the key role of the Dukhin number in controlling  the local channel selectivity.
Eq.~(\ref{eq:selectivities}) quantifies the relative importance of the counterion flux over the total transport. Due to the conical shape of the channel, $\gamma_+^R$ is larger than $ \gamma_-^L$, meaning that counterion transfer  in presence of an external driving is larger at the tip than at the base.  Such imbalance in selectivities results in a transient ion readjustment  when an external driving is switched on. In the case of counterions moving from tip to base (negative $\Delta V$) this imbalance in selectivities results in a transient accumulation of ions inside the channel. On the contrary, when counterions move from base to tip (positive $\Delta V$) there will be a relative larger amount of ions leaving than entering the channel resulting in an overall decrease of salt concentration.  In either case,  the stationary state is reached when the nonequilibrium accumulation/depletion dynamics counterbalances the asymmetry of local selectivity induced by the geometry.
Eq.~(\ref{eq:selectivities}) implies that the imbalance in selectivities is controlled by the asymmetry between $Du/h_L$ and $Du/h_R$. Both  $Du \ll 1$ and $Du \gg 1$ result in a uniform selectivity between the two ends of the channel, \textit{i.e.} no rectification (see Fig.~(\ref{FIG:selectiv})). High Dukhin number, $Du \gg 1$, means that the selectivity of counterions at either end tends to one (that is the selectivity of coions tends to zero): the coions are completely excluded from the system and the geometrical asymmetry is nullified by the perfect selectivity of the channel. No bulk transport is present so that the entirety of transport takes place in the EDL. On the other side, for $Du \ll 1$ the selectivity at either ends tends to its bulk value $1/2$. In this regime, irrespectively of the physical extension of the EDL the entirely of transport takes place in the unselective bulk and the omhic bulk response is restored. 
\\
The  asymmetry between $Du^L/h_L$ and $Du^R/h_R$ is maximized for $Du \sim 1$ (in our case $Du \sim 1/2$ because of the normalization used for the marginal concentrations). \\ 
The qualitative interpretation of ICR  caused by an asymmetry in the local selectivity at either end of the nanochannel is qualitatively consistent with the  pioneer proposal of  Woarmann~\cite{B301021J}. However, our analysis  provides a fresh interpretation of an old puzzle. We have shown  that $Du$ is the principal electrostatic parameter that locally controls the channel selectivity, with a secondary effect due to $\lambda_D$, while  Woermann pointed at $\lambda_D$ as the main length to be compared with  channel confinement.  
Although it may fly against intuition,  it is not the physical size of the EDL that determines the system capability to rectify ionic current.

\section{Conclusion}
In summary, we have presented here a theoretical analysis to address the phenomenon of ionic current rectification in nanometric channels. We have specifically focused on the case of a geometric ionic diode where the symmetry breaking is caused only by the conical geometry of the system. The theoretical framework mainly relies on two assumptions: a slowly varying channel geometry and a small electrostatic potential variation in the transversal direction. These ingredients allow us to derive formal expressions for the electrostatic potential, Eq.~(\ref{eq:potential}), and for the ionic current, Eq.~(\ref{eq:current}), and to explore the response of the system for different values of $\lambda_D$ and $Du$. The main outcome of the work is the identification of the Dukhin length as the primary electrostatic length scale controlling rectification. It follows that rectification is expected to be measured in systems with size comparable to the Dukhin length, which remarkably can reach the micrometer scale~\cite{B909366B}. This fact may explain recent experimental works\cite{doi:10.1021/jacs.6b11696, doi:10.1021/acs.jpclett.7b03099} in which ICR is observed in mesoscopic pores. 
\\ To conclude by misquoting Wolfrang Pauli \footnote{as Pauli once said \textit{God makes the bulk; the surface was invented by the devil}}, it is a dynamical usage of surfaces that let the nanofluidic diode succeed where demons don't.

\begin{acknowledgments}
S.D.L acknowledges enlightening discussions with A. R. Poggioli. I.P. acknowledges support from MINECO under project FIS2015-67837-P and Generalitat de Catalunya under project 2017SGR-884 and SNF Project No. 20021-175719. The work has been funded by the European Union's Horizon 2020 research and innovation program under ETN grant  674979-NANOTRANS.
\end{acknowledgments}
\section{Appendix}
Here we report some details of the implementation in COMSOL for the numerical integration of the following equations:
\begin{eqnarray}\label{eq:thesystem}
J_+ &=& -\partial_x c_+ + c_+ [\partial_x \log h -(\partial_x \langle \phi \rangle - \Delta V ) +\partial_x \log \langle e^{-\psi} \rangle ] \nonumber \\
J_- &=& -\partial_x c_- + c_- [\partial_x \log h +(\partial_x \langle \phi \rangle - \Delta V ) +\partial_x \log \langle e^{+\psi} \rangle ] \nonumber \\
\partial_y^2 \psi &=& -\frac{\lambda_D^{-2}}{2 h} [(c_+ - c_-) - (c_+ + c_-) \psi]  
\end{eqnarray}
First of all let us recall here the appropriate boundary conditions  for the system at hands. In both ends of the channel we have to impose continuity in the electrochemical potential for each species. Starting from the left side we write in adimensional variables:
\begin{equation}\label{eq:continuitychemicalpotential}
\log \frac{1}{2} \pm \frac{\Delta V}{2} = \log c_{\pm} (0) + \beta A_{\pm}(0)
\end{equation}
where by definition:
\begin{equation}\label{eq:defA}
e^{-\beta A_{\pm}(0)} = e^{\mp \frac{\Delta V}{2}} \int_{-h_L}^{+h_L} dy\ e^{\mp \phi (0,y)}
\end{equation}
By substituting eq.(\ref{eq:defA}) into (\ref{eq:continuitychemicalpotential}) we obtain:
\begin{equation}\label{eq:boundaryczero}
c_{\pm} (0) = \frac{1}{2} \int_{-h_L}^{+h_L} dy\ e^{\mp \phi_L(y)}
\end{equation}
where the boundary condition for $c_{\pm}(0)$ is expressed in terms of the function $\phi_L(y)  \equiv \phi(0,y) $. The latter is obtained by solving the following transversal equation at $x=0$:
\begin{widetext}
\begin{equation}\label{eq:boundaryleft}
\partial_y^2 \phi_L = - \frac{\lambda_D^2}{2 + \kappa} \left[  \int_{-h_L}^{+h_L} \sinh (\phi_L) - (\phi_L - \langle \phi_L \rangle) \int_{-h_L}^{+h_L} \cosh (\phi_L) \right], \hspace*{1cm}{x=0}
\end{equation}
\end{widetext}
where $\langle \phi_L \rangle = \frac{1}{2 h_L} \int_{0}^{h_L} dy \phi_L(y)$ , $h_L = 1+ \frac{\kappa}{2}$ and we made use of the fact that:
\begin{eqnarray}
c_+(0) + c_-(0) = \int_{-h_L}^{+h_L} dy \cosh(\phi_L(y)) \nonumber \\
c_+(0) - c_-(0) = \int_{-h_L}^{+h_L} dy \sinh(\phi_L(y))
\end{eqnarray}
Eq. (\ref{eq:boundaryleft}) can be then numerically integrated using the standard electrostatic boundary conditions:
\begin{eqnarray}
\partial_y \phi_L(0) &=&0 \nonumber \\
\partial_y \phi_L(\pm h_L) &=& \mp \frac{Du}{\lambda_D^2}
\end{eqnarray}
Likewise we find the appropriate boundary value for $c_{\pm}(1)$ using :
\begin{equation}\label{eq:boundarycone}
c_{\pm}(1) = \frac{1}{2} \int_{-h_R}^{+h_R} dy\ e^{\mp \phi_R}
\end{equation}
Therefore the expressions (\ref{eq:boundaryczero}) and (\ref{eq:boundarycone}) are now numbers which can be directly used as boundary conditions for the system in (\ref{eq:thesystem}).
\\
It is also convenient in COMSOL to rescale the $y$ variable in the following way:
\begin{eqnarray}
&&y \to h(x) y\prime \\
&&f(x,y) \to f(x, h(x)y\prime) \equiv f\prime(x, y\prime)
\end{eqnarray}
In this way we map the original domain to a square domain substantially simplifying the COMSOL calculation.
From the chain rule it follows:
\begin{widetext}
\begin{subequations}\label{eq:chainrule}
\begin{eqnarray}
\partial_x f(x, y) \to \partial_x f\prime(x, y\prime) &=&\partial_x f\prime(x, y\prime) + \partial_{y\prime} f\prime(x, y\prime) \partial_x y\prime \nonumber \\ &=&  \partial_x f\prime(x, y\prime) - \partial_{y\prime} f\prime(x, y\prime) \frac{y\prime}{h(x)} d_x h(x) \\
\partial_y f(x,y) \to \partial_y f\prime(x, y\prime) &=& \partial_{y\prime} f\prime(x, y\prime) d_y y\prime \nonumber \\ &=& \partial_{y\prime} f\prime(x, y\prime) \frac{1}{h(x)}
\end{eqnarray}
\end{subequations}
\end{widetext}
The only variables in the model that depend on $y$ are $\psi (x,y)$, $\phi_L(y)$ and $\phi_R(y)$. For each of them we apply (\ref{eq:chainrule}) so that the electrostatic boundary condition for $\psi$ (likewise for $\phi_L$ and $\phi_R$) become:
\begin{eqnarray}
\partial_{y\prime} \psi\prime (x, y\prime=0) &=& 0 \\
\partial_{y\prime }\psi\prime (x, y\prime =\pm 1) &=& \mp \frac{Du}{\lambda_D^2} h(x)
\end{eqnarray}
and the rescaled Poisson equation:
\begin{widetext}
\begin{equation}
\partial^2_{y\prime} \psi\prime = -\frac{ \lambda_D^{-2}}{2} h(x)\left[ (c_+ - c_-) - \psi\prime (c_+ + c_-) \right]
\end{equation}
\end{widetext}


%
%

%


\newpage

\bibliography{draft_diode}

\end{document}